# Characterization of nanoscale mechanical heterogeneity in a metallic glass by dynamic force microscopy

Y.H. Liu [1], D. Wang [1], K. Nakajima [1], W. Zhang [2], A. Hirata [1], T. Nishi [1], A. Inoue [2] and M.W. Chen [1]

[1] *WPI-Advanced Institute for Materials Research, Tohoku University, Sendai 980-8577, Japan*
[2] *Institute for Materials Research, Tohoku University, Sendai 980-8577, Japan*

[*] To whom correspondence should be addressed. E-mail: mwchen@wpi-aimr.tohoku.ac.jp

## Abstract

We report nano-scale mechanical heterogeneity of a metallic glass characterized by dynamic force microscopy. Apparent energy dissipation with the variation of ~12%, originating from non-uniform distribution of local viscoelasticity, was characterized. The correlation length of heterogeneous viscoelasticity was measured to be ~2.5±0.3 nm, which is well consistent with the dimension of shear transformation zones for plastic flow. This study provides the first experimental observation on the nano-scale mechanical heterogeneity in a metallic glass, and may fill the gap between atomic models and the macroscopic properties of metallic glasses.

Metallic glasses are vitrified solids quenched from liquids through glass transition and inherit the disordered structure of liquids with intrinsic topological and geometric frustrations as well as a large number of quench-in defects [1-5]. The lack of a long-range atomic periodicity of metallic glasses leads to the nano-scale heterogeneity in the distribution of the inherent defects, giving rise to the formation of densely packed atomic clusters and loosely packed defective domains. Since the constituent atoms in defective domains have lower atomic coordination than those in the dense atomic clusters, inelastic and anelastic relaxation becomes possible by local atom rearrangements. These sites have thus been suggested as the preferred regions that initiate the glassy structure destabilization caused by either high temperatures or applied shear stresses, which play a crucial role in the mechanical behavior and glass transition of metallic glasses [5-12]. Although extensive efforts have been devoted to elucidating the structural and mechanical heterogeneity [8-17], direct observations of the nano-scale mechanical heterogeneity are still missing and the nature of the heterogeneity remains poorly known.

Amplitude modulation dynamic atomic force microscopy (AM-AFM), with vibrating cantilever-tip ensemble scanning across a sample, is a powerful tool to characterize nano-scale material properties by measuring the phase shift, $\phi$, arising from the energy dissipated, $E_{\text{dis}}$, during tip-sample interactions:

$$E_{dis} = (\sin\phi - \frac{\omega}{\omega_0} \cdot \frac{A}{A_0}) \cdot \frac{\pi \cdot k \cdot A \cdot A_0}{Q} \qquad (1)$$

in which $\omega$ is the drive frequency, $\omega_0$ is the resonant frequency of cantilever, $A$ is the vibration amplitude during testing, $A_0$ is the free amplitude without tip-sample interaction, $k$ is the spring constant of cantilever, and $Q$ is the quantity factor [18-23]. The technique has been extensively employed to discriminate heterogeneous structures in a variety of materials [19-22], but its application on metallic glasses has been limited by making a damage-free sample surface with sub-nano-scale roughness. In the present work, we investigated the nano-scale mechanical behavior of an atomically flat and damage-free metallic glass film using AM-AFM and provides the first experimental evidence on the nano-scale mechanical heterogeneity.

A 2-μm-thick metallic glass film was prepared by rf-magnetron sputtering. A $Zr_{55}Cu_{30}Ni_5Al_{10}$ metallic glass was used as the sputtering target [25]. **Figure 1(a)** shows the X-ray diffraction (XRD) spectra of the as-deposited film and the target. Except minor peak broadening of the film, two spectra are very analogous to each other and both free of crystalline peaks. Thermal analysis by differential scanning calorimeter (DSC) further confirms the glassy state of the film with an evident glass transition (**Fig. 1(b)**). The slightly lower glass transition temperature ($T_g$) of the film than that of the bulk target may be due to the much higher cooling rates of the film deposition. The lower crystallization temperature appears to be caused by the large surface of the film which can provide preferred nucleation sites for crystallization. The microstructure of the film was inspected by a Cs-corrected transmission electron microscope (TEM). The micrographs taken by scanning TEM (STEM) and high-resolution TEM (HRTEM) demonstrate the amorphous structure of the film (**Fig. 1(c)** and **(d)**). Slight variation in bright/dark contrast with a scale of 2-3 nm in the STEM image (**Fig. 1(d)**) may correspond to the structural heterogeneity of the thin film since detectable composition difference between the dark and bright domains cannot be found.

The as-deposited film was directly used for the AM-AFM measurement without any further polishing and cleaning to avoid possible damage and contamination (**Fig. S1**). AM-AFM scanning was performed with a scanning probe microscope (Multi-mode with a Nanoscope V controller) operated at near the resonant frequency (~ 150 kHz) of the Si cantilever. A sharp diamond tip with a ~1 nm apex (*see* **Fig. S2**) was used for a high spatial resolution [23]. The stress between the sharp tip and the sample is designed to be smaller than the yield stress of the metallic glass, and only elastic deformation is involved into the measurement (**Fig. S3**). Surface topography, phase shift, and amplitude images were recorded simultaneously during the scanning (**Fig. 2(a)** and **(b)**). The topographic profile depicted in **Fig. 2(c)** demonstrates that the height variation is less than 0.9 nm, only about 2-3 atomic layers on average. **Fig. 2(b)** shows the phase shift image of the metallic glass in which the evident phase contrast indicates the nanoscale variation in the phase shift. The phase shift profile in **Fig. 2(c)** reveals that the discrepancy in the phase shift is as large as ~6°. Direct comparison between the phase shift and height images demonstrates that the phase shift contrast does not have visible correlation with the surface topography. The height and phase shift profiles (**Fig. 2(c)**) taken along the same line also demonstrate that the phase shift is independent of the surface roughness since the regions with large phase shift do not correspond to rough domains. This is also verified by the standard sample of atomically flat graphite. (*see* **Fig. S4**). Therefore, the influence of the topography on the observed phase shift of the metallic glass film can be rationally ignored and the contrast in the phase shift image mainly reflects intrinsic material characteristics. Additionally, the phase shift shown in **Fig. 2(b)** can be reproduced from the films with different thicknesses (**Fig. S5**), indicating that the observed heterogeneity in the phase shift is independent of residual stresses that may be significant in thin films and vary with film thickness [26].

Viscoelasticity of the metallic glass and surface energy hysteresis are two possible origins of the measured phase shift by AM-AFM [19]. However, their contributions can be readily discriminated by measuring the normalized energy dissipation $E^*_{dis}$ curve, $E^*_{dis}=E_{dis}/E^{max}_{dis}$ with $E^{max}_{dis}$ the maximum of the $E_{dis}$ vs. $A/A_0$ curve, that is independent of experimental parameters [19, 22]. According to the tip-sample interaction, either viscoelasticity or surface energy hysteresis has its own unique features in the $E^*_{dis}$ vs. $A/A_0$ and $\delta E^*_{dis}/\delta(A/A_0)$ vs. $A/A_0$ curves. For materials exhibiting viscoelastic behavior, the tip-sample interaction relies on both deformation and the deformation rate, giving rise to the dissipation inflection at the end of the $A/A_0$ range, and thereby the $\delta E^*_{dis}/\delta(A/A_0)$ vs. $A/A_0$ curve is featured by an extreme point [19]. In this study, the average $E^*_{dis}$ vs. $A/A_0$ and $\delta E^*_{dis}/\delta(A/A_0)$ vs. $A/A_0$ curves of the metallic glass (**Fig. 2(d)-(e)**) exhibit the typical characteristic of viscoelasticity, demonstrating that the observed phase shift in **Fig. 2(b)** mainly originates from the viscoelastic behavior of the material. The viscosity difference between the high and low phase shift regions is ~10%, calculated according to the Eq. (6)

of Ref. [19]. The spatially uneven viscosity is most likely associated with the heterogeneity of elastic modulus, as suggested by recent XRD experiment [17].

**Fig. 3(a)** is the energy dissipation map converted from the phase shift and amplitude images according to **Eq. (1),** in which the nano-scale heterogeneity in the energy dissipation is evident. Again, the energy dissipation distribution does not show any correlation with the surface roughness as evidenced by the significant difference between the energy dissipation map and the differential of the height image that highlights the effect of surface steps (**Fig. 3(b)**). The statistic distribution of the energy dissipation can be perfectly fitted by the Gaussian distribution (**Fig. 3(c)**), indicating the viscoelastic heterogeneity in the metallic glass is random and the measurement does not introduce any systematic error. Although the mean energy dissipations $\overline{E}_{dis}$ depends on the $A/A_0$ ratios, the $\overline{E}_{dis}$ at each $A/A_0$ ratio can be scaled by the corresponding full-width half-maximum (FWHM), $E_{dis}^{FWHM}$ to a constant of ~12% for the averaged variation of the energy dissipation caused by the viscoelastic heterogeneity (**Fig. 3(d)**), in consistence with the estimated average viscosity difference in the low and high phase shift domains. This further demonstrates that the local heterogeneity revealed by the AM-AFM measurement is an inherent characteristic of the metallic glass. The ~2 nm domains with the large energy dissipation above the average show significant anelastic deformation when the tip strikes the sample surface, which are believed to correspond to the more defective regions that have more "liquid-like" behavior [9, 12] because of their relatively low viscosity and elastic modulus.

The characteristic lengths of the viscoelastic heterogeneity and the surface roughness are evaluated by the correlation functions $P(r)=<P(r)-P(0)>^2$ and $H(r)=<H(r)-H(0)>^2$, in which $P(r)$ and $H(r)$, and $P(0)$ and $H(0)$ are the phase shift and height at the coordinate (x, y) and the reference position ($x_0$, $y_0$), respectively [27]. The calculated correlation functions for both phase shift and surface roughness are plotted in **Fig. 4(a)**, which follow $P(r)=2\sigma^2[1-\exp(-(r/\xi)^{2\alpha}]$ and $H(r)=2\sigma^2[1-\exp(-(r/\xi)^{2\alpha}]$, where $\sigma$ is the root mean square roughness or phase shift, $\alpha$ is the roughness (or phase shift) exponent, and the lateral correlation length $\xi$ defines the distance between two correlated points [27]. Based on the data fitting, it can be found that the correlation length for the phase shift is ~2.5 nm whereas the length for the surface roughness is ~9 nm (**Fig. 4(a)**). Moreover, these values are independent of the $A/A_0$ ratios (see **Fig. 4(b) and (c)**), indicating that the correlation lengths are a material characteristic. The significant difference in the correlation lengths between the phase shift and surface roughness further validates that the heterogeneous phase shift measured by AM-AFM comes from the intrinsic material behavior.

It is interesting to note that the correlation length of the visocelastic heterogeneity is in good agreement with the characteristic length (2-3 nm) of the low size limit of shear transformation zones (STZs) for plastic flow in Zr-based metallic glasses [28, 29, 30], implying that they may share the same physical origin. Based on MD simulations, both dynamic heterogeneity and STZs are known to associate with the uneven atomic arrangements in metallic glasses, particularly, defective regions constituted by loosely packed atoms [5, 11, 28, 29, 31, 32]. Thus, heterogeneous viscoelasticity appears to arise from the atomic structural heterogeneity, in which densely and loosely packed regions represent different viscoelastic behavior because of the variation in time-related atomic mobility. Moreover, the wide range of energy dissipation indicates that there is a broad distribution of energy barriers for shear transformation and structure relaxation. Upon either mechanical loading or thermal heating, the structure changes of metallic glasses do not happen homogeneously, but starts preferentially from the some regions where the viscoelastic behavior is more substantial because of their high capability to absorb and convert the input energies for the formation of STZs or local glass transition [33].

Since the film was prepared by sputtering with a high cooling rate [35], the constitute atoms have less time to reach the equilibrium packing state. The as-deposited film may has a much looser atomic packing than the slowly quenched bulk counterpart [36], which may lead to more prominently heterogeneous behavior. Upon annealing, the film can be relaxed towards equilibrium by annihilating out the excess quench-in defects for a very similar structure to the bulk sample [36]. Indeed, our measurements on the annealed film (**Fig. S6**) show the reduced variation of phase shift of ~8% and the increased correlation length of ~4.2 nm that is close to the upper size limit of the STZs in Zr-based BMGs [30]. The increased correlation length actually coincides with the recent simulations in which densely packed glasses by slow quenching enhances the percolation of short range order [11, 12, 37].

In summary, we experimentally characterized the nano-scale mechanical heterogeneity of a metallic glass by taking the advantage of dynamic AFM. The measured correlation lengths are comparable to the sizes of STZs and the characteristic length of secondary relaxation in metallic glasses, which provides important insights on the instability of metallic glasses subjected to applied stresses and high temperatures and has important implications in understanding the atomic mechanisms of mechanical properties of metallic glasses.

**Figure Captions**

**FIG. 1** (a) XRD spectra of the glassy film and bulk metallic glass. (b) DSC traces of the film together with the bulk samples. (c) HREM image of the film showing uniform maze-like pattern. (d) STEM image shows some bright/dark contrast that may correspond to the structural heterogeneity of the glassy sample.

**FIG. 2** (a) The height image with rms roughness of ~0.3 nm; and (b) the phase shift image. (c) Height and phase shift profiles taken from the same region. (d) The normalized energy dissipation $E^*_{dis}=E_{dis}/E^{max}_{dis}$; and (e) its derivative as function of the amplitude ratio $A/A_0$.

**FIG. 3** (a) Energy dissipation map with the amplitude ratio $A/A_0=0.90$. (b) Differential of height image which shows distinctly different morphology from the energy dissipation map. (c) The distribution of energy dissipation. The solid line is the Gaussian fit of the experimental data points. (c) Plot of $E^{FWHM}_{dis}/\overline{E}_{dis}$ as function of the amplitude ratio $A/A_0$. The average difference in viscoelasticity is estimated to be ~12%.

**FIG. 4** (a) Estimations of correlation lengths for both height and phase shift. (b) and (c) Plots of correlation lengths of phase shift and height images as function of the amplitude ratio $A/A_0$. For different $A/A_0$ ratios, the correlation length in the phase shift and height images is 2.5±0.3 nm and 7.3±1.5 nm, respectively.

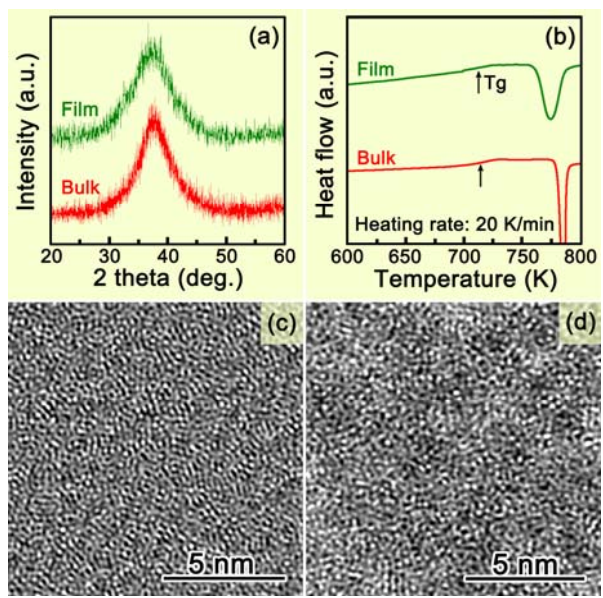

**Fig. 1** Liu *et al*.

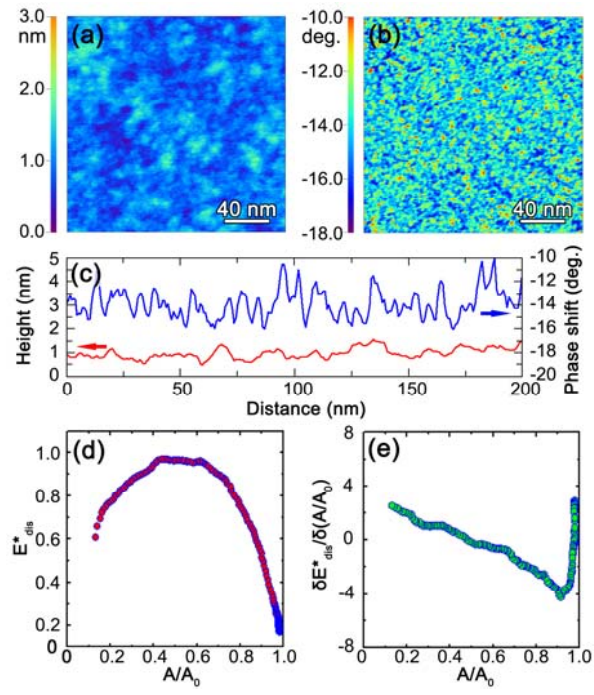

**Fig. 2** Liu *et al*.

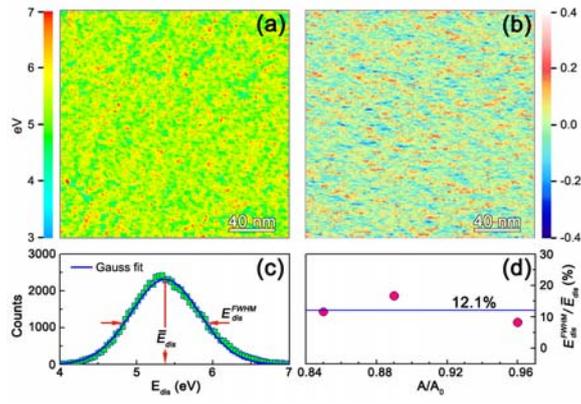

**Fig. 3** Liu *et al*.

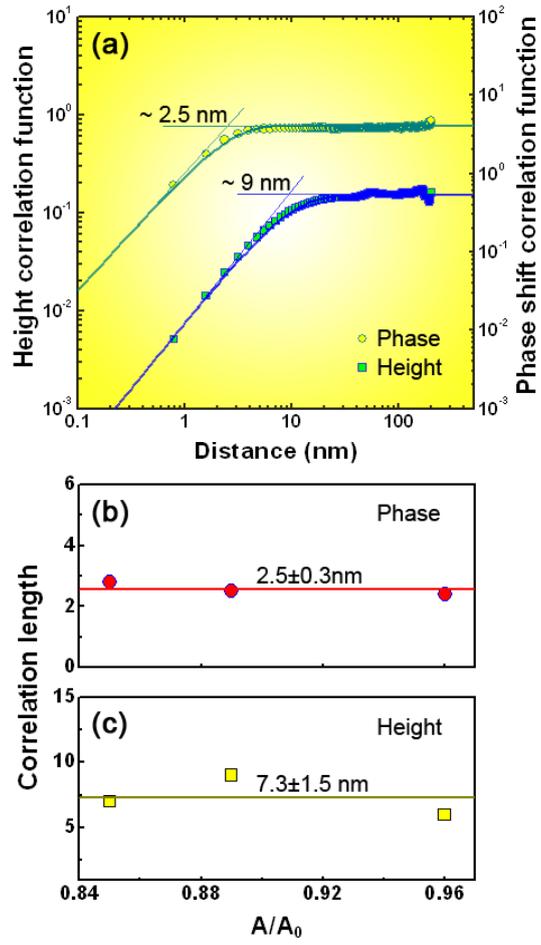

**Fig. 4** Liu *et al*.